\documentclass[12pt]{iopart}
\usepackage{graphicx}

\newcommand{\be}{\begin{equation}}
\newcommand{\ee}{\end{equation}}
\newcommand{\bea}{\begin{eqnarray}}
\newcommand{\eea}{\end{eqnarray}}

\begin{document}

\title{Curie and N\'eel Temperatures of Quantum Magnets}

\author{J. Oitmaa and Weihong Zheng
}

\address{School of Physics, The University of New South Wales, Sydney, NSW 2052, Australia}

\begin{abstract}
We estimate, using high-temperature series expansions, the transition temperatures
of the spin $\frac{1}{2}$, 1 and $\frac{3}{2}$ Heisenberg ferromagnet and antiferromagnet in 3-dimensions.
The manner in which
the difference between Curie and N\'eel temperatures vanishes with increasing
spin quantum number is investigated.
\end{abstract}



\maketitle

It is well known that in classical spin models, such
as as the Ising or classical Heisenberg models, on bipartite lattices the critical temperature
(if it exists) is the same for ferromagnetic exchange (Curie temperature)
as for antiferromagnetic exchange (N\'eel temperature). This is a direct
consequence of the free energy being an even function of the exchange parameter $J$.
It has also been known for some time, but perhaps less widely, that for the
quantum spin-$\frac{1}{2}$ Heisenberg model the Curie and N\'eel temperatures
are unequal. Early work \cite{Rus63,Rushbrooke74} put the N\'eel temperature some 10\%
above the Curie temperature for spin-$\frac{1}{2}$, for both the simple cubic
(SC) and body-centred cubic (BCC) lattices, with the difference decreasing
rapidly with increasing $S$. However these results were based on rather short
series (six terms) and the critical point estimates contained large uncertainties.

We have re-investigated this question, using substantially longer series
(14th order for $S=\frac{1}{2}$, 12th order for $S=1$, 9th order for $S=\frac{3}{2}$). This is made possible
not only by the massive increase in computing power now available, but also
by the development of efficient linked-cluster expansion methods.
The reader is referred to a recent review \cite{Gelfand00:49}
for further details of this method.

The Hamiltonian is written in the form
\be
H = - J \sum_{\langle ij\rangle } {\bf S}_i \cdot {\bf S}_j
 - h \sum_i S_i^z - h_s \sum_i \eta_i S_i^z \label{eq_H}
\ee
where the ${\bf S}_i$ are spin-$S$ operators, $h$ and $h_s$ are uniform and staggered fields,
with $\eta_i = \pm 1$ on respective sublattices, and the interaction is taken between
nearest neighbours $\langle ij\rangle$. $J>0$ ($<0$)
corresponds to the ferromagnet (antiferromagnet). While (\ref{eq_H})
contains the form of the exchange energy for a real spin-$S$ system, for comparison
between different $S$-values and, in particular, for passage to the classical
limit $S\to \infty$, it is convenient to write $\tilde{J} = J S (S+1)$ and to express
critical temperatures in units of $\tilde{J}/k_B$.

The critical temperature $k_B T_c/J$ is most reliably obtained from the strongly
divergent `ordering' susceptibility in zero field: the uniform susceptibility $\chi$
for the ferromagnet or the staggered susceptibility $\chi_s$ for the antiferromagnet.
High-temperature series for these quantities can be derived in the form
\be
\chi, \chi_s =  \sum_{r=0}^{\infty} a_r K^r \label{eq_chi_ser}
\ee
where $K = \vert J \vert/k_B T$ and the $a_r$ are numerical coefficients. The uniform
susceptibility for the spin-$\frac{1}{2}$ case is  known through order
$K^{14}$ \cite{Oitmaa96:53}, for both the SC and BCC lattices. In the present paper
we give the staggered susceptibility series to the same order. This represents an
addition of six new terms to the previously known series\cite{Pan99}.
At the same time we compute uniform and staggered susceptibilities for the
$S=1$ case, through order $K^{12}$ and $K^{11}$ respectively, for both lattices,
extending the previous series by five terms.
We have also calculated the corresponding series for $S=\frac{3}{2}$ through order $K^9$.
For the classical $S=\infty$ model
the susceptibility series is known through $K^{21}$\cite{Butera97} and we will use this
series in our comparison.

Tables \ref{tab1}, \ref{tab2} and \ref{tab2B} list the series coefficients, in integer format, for both the SC and
BCC lattices. The coefficients are positive and appear to be quite regular, suggesting
that the radius of convergence is determined by the physical singularity on the
positive real axis (we will return to this point later!) Closer inspection, however,
reveals some oscillation, reflecting interference from non-physical singularities
near the circle of convergence. Although we do not base our analysis on this, it is
instructive to see a ratio plot\cite{Guttmann89}. We show such a plot in Figure 1.
Looking at the SC lattice first, it is evident that the $S=\frac{1}{2}$ series, in
particular, shows a strong 4-term oscillation. This results from a pair
of singularities on, or near, the imaginary axis, near the circle of convergence.
The $S=1$ series are much  more regular and, qualitatively, look quite similar
to the $S=\infty$ case. The BCC series are rather regular, even for $S=\frac{1}{2}$. There is a 2-term
oscillation in all series, which is characteristic of bipartite lattices.
Apart from the $S=\frac{1}{2}$ (SC) case, resonable estimates of the
Curie and N\'eel temperatures can be made visually. Unless something totally
unexpected were to occur at higher orders, it seems clear that the N\'eel
temperature exceeds the Curie temperature for both $S=\frac{1}{2}$ and $S=1$
(remembering that the intercept on the ordinate axis is $k_BT_c/\tilde{J}$).
The very similar limiting slope of the different plots is consistent with the
universality expectation that all quantities diverge with the same exponent $\gamma$.

To obtain more accurate estimates of the critical parameters we turn to
Pad\'e approximants\cite{Guttmann89}. Tables \ref{tab3}, \ref{tab4}
give estimates of the critical point $K_C$ and exponent $\gamma$, assuming
a normal power-law singularity
\be
\chi, \chi_s \sim C_0 (1-K/K_C)^{-\gamma}; \quad K \to K_C- \label{eq_cri}
\ee
obtained from high-order Pad\'e approximants to the logarithmic derivative series.
Different approximants give quite consistent results and we summarize the overall
estimates of the critical temperature in Table \ref{tab5}. The exponent estimates
from the highest approximants are around 1.42 ($S=\frac{1}{2}$), 1.41 ($S=1$).
Early studies of the $S=\infty$ series also gave values in this range, although
the recent long series give lower values, approaching the field theory
prediction $\gamma\simeq 1.39$. Our results are consistent with
the universality expectation.

Figure 2 shows plots of critical temperatures $k_B T_c/S(S+1)J $ versus
$1/S(S+1)$. The plots appear linear, particularly if the $S=\frac{1}{2}$
points are excluded, and indicate that, to a very good approximation
\be
k_B T_c /J \sim a S (S+1) + b
\ee
where $a$, $b$ are constants independent of $S$. Their values are

\begin{center}
\begin{tabular}{ccccc}
  & \multicolumn{2}{c}{SC}  & \multicolumn{2}{c}{BCC} \\
  & $\chi$ & $\chi_s$ & $\chi$ & $\chi_s$ \\
$a$ &  \multicolumn{2}{c}{1.4429} & \multicolumn{2}{c}{2.0542} \\
$b$ & $-$0.288 & $-$0.150 & $-$0.320 & $-$0.174 \\
\end{tabular}
\end{center}
This linear relation may then be used to obtain reliable estimates
of Curie and N\'eel temperature for values $S>\frac{3}{2}$.


With our longer series we are also able, for the first time, to estimate
values for the amplitudes $C_0$ of the leading singular term (\ref{eq_cri}).
This is done in two ways. The first is to use the estimates of $K_C$, $\gamma$
obtained previously, form the series for
\be
(1-K/K_C)^{\gamma} \chi \sim C_0 + \cdots; \quad K \to K_C-
\ee
and evaluate Pad\'e approximants to this series at $K_C$. The second is to
compute the series for
\be
\chi^{1/\gamma} = C_0^{1/\gamma} (1-K/K_C)^{-1}
\ee
Pad\'e approximants to this series should have a simple pole at $K_C$ with
residue $K_C C_0^{1/\gamma}$. Both methods give consistent results. We give in
Table \ref{tab5} our best estimates and error. As usual with series analysis, these
are not true statistical errors but only confidence limits based on the spread
of results. As can be seen from Table \ref{tab5}, these amplitudes are all of order 1
and show a decrease of some 30\% on going
from $S=\frac{1}{2}$ to $S=\infty$, with the antiferromagnetic amplitude some
5\% smaller than the ferromagnetic one.

The conclusion that the Curie temperature $T_C$ is lower than the N\'eel
temperature $T_N$ has a puzzling consequence, as has been remarked on before
\cite{Rushbrooke74}. Assuming that the ferromagnetic susceptibility $\chi(K)$
also has a weak, energy like singularity at $-K_N$ ($K_N<K_C$), as is known
to be the case for the Ising model, means that the radius of convergence
of the series is $\vert K_N\vert $. Hence the series coefficients must,
at some point, begin to alternate in sign.
To check this point further we follow the procedure of Baker {\it et al.}\cite{Baker67},
in seeking evidence for a singularity at $-K_N$
in the uniform susceptibility, and at $-K_C$ in the staggered susceptibility. To this
end we form the series for

\be
F(K) = {d\over dK} {\Big (} {d\over dK} \ln \chi (K) - {\gamma\over K_C-K} {\Big )} \label{eq5}
\ee
and
\be
F_s(K) = {d\over dK} {\Big (} {d\over dK} \ln \chi_s (K) - {\gamma\over K_N-K} {\Big )} \label{eq6}
\ee
The first step substracts out the dominant physical singularity from the logarithmic
derivative series. This series is expected to have a weak singularity at the corresponding
N\'eel or Curie point. The final differentiation is to strengthen this singularity
In Table \ref{tab6} we show estimates of the location of this secondary singularity
and the corresponding residue for the $S=\frac{1}{2}$ series on the BCC lattice. As is clear, the series $F(K)$
shows a consistent pole at $K\simeq -0.72$, consistent with the direct estimate
of $K_N$ (Table \ref{tab3}).
Similarly the series $F_s(K)$ shows a consistent pole at $K\simeq -0.799$,
consistent with the direct estimate of $K_C$ (Table \ref{tab3}).
These numerical estimates will, of course, depend on the choice made for
$K_C$, $K_N$, $\gamma$ in Eqs. (\ref{eq5}) and (\ref{eq6}), but are found to be
relatively insensitive to this choice. We have not repeated
this analysis for the SC case or for $S=1$, $\frac{3}{2}$.

\begin{table}
\caption{\label{tab1}Series for $\chi$ and $\chi_s$ for spin-$\frac{1}{2}$. To avoid fractions a multiplier
$4^{n+1} n!$ for $\chi$, or $4^{n+1} (n+1)!$ for $\chi_s$ has been used, where $n$ is the power of $K$.}
\begin{indented}
\item[]\begin{tabular}{@{}rrr}
\br
n & $\chi$ & $\chi_s$ \\
\mr
\multicolumn{3}{c}{Simple Cubic Lattice $S=\frac{1}{2}$} \\
 0 &                   1 &                        1 \\
 1 &                   6 &                       12 \\
 2 &                  48 &                      168 \\
 3 &                 528 &                     2880 \\
 4 &                7920 &                    59376 \\
 5 &              149856 &                  1478592 \\
 6 &             3169248 &                 42357024 \\
 7 &            77046528 &               1353271296 \\
 8 &          2231209728 &              48089027328 \\
 9 &         71938507776 &            1908863705088 \\
10 &       2446325534208 &           83357870602752 \\
11 &      92886269386752 &         3926123179720704 \\
12 &    3995799894239232 &       198436560561973248 \\
13 &  180512165153832960 &     10823888709015846912 \\
14 & 8443006907441565696 &    635114442481347244032 \\
\mr
\multicolumn{3}{c}{Body Centred Cubic Lattice $S=\frac{1}{2}$} \\
 0  &                       1 &                         1 \\
 1  &                       8 &                        16 \\
 2  &                      96 &                       320 \\
 3  &                    1664 &                      8192 \\
 4  &                   36800 &                    248768 \\
 5  &                 1008768 &                   8919296 \\
 6  &                32626560 &                 367854720 \\
 7  &              1221399040 &               17216475136 \\
 8  &             51734584320 &              899434884096 \\
 9  &           2459086364672 &            51925815320576 \\
10  &         129082499311616 &          3280345760086016 \\
11  &        7432690738003968 &        225270705859919872 \\
12  &      464885622793134080 &      16704037174526894080 \\
13  &    31456185663820136448 &    1330557135528577925120 \\
14  &  2284815238218471260160 &  113282648639921512955904 \\
\br
\end{tabular}
\end{indented}
\end{table}

\begin{table}
\caption{\label{tab2B}Series for $\chi$ and $\chi_s$ for spin-1. To avoid fractions a multiplier
$3^{n+1}n!/2$ ($3^{n+1}(n+1)!/2$) has been used for $\chi$ ($\chi_s$) series, where
$n$ is the power of $K$.}
\begin{indented}
\item[]\begin{tabular}{@{}rrr}
\br
n & $\chi$ & $\chi_s$ \\
\mr
\multicolumn{3}{c}{Simple Cubic Lattice $S=1$}  \\
 0  &                    1 &                     1  \\
 1  &                   12 &                    24  \\
 2  &                  222 &                   702  \\
 3  &                 5904 &                 26280  \\
 4  &               201870 &               1184526  \\
 5  &              8556912 &              63357984  \\
 6  &            426905802 &            3887604666  \\
 7  &          24674144724 &          270348199128  \\
 8  &        1616505223518 &        20988390679758  \\
 9  &      118701556096392 &      1802403961243776  \\
10  &     9628527879611262 &    169418364565523958  \\
11  &   856813238084411136 &  17314303199655636792  \\
12  & 82856991914713902402 &                        \\
\mr
\multicolumn{3}{c}{Body Centred Cubic Lattice $S=1$}  \\
 0  &                      1 &                     1 \\
 1  &                     16 &                    32 \\
 2  &                    424 &                  1320 \\
 3  &                  16512 &                 71136 \\
 4  &                 819240 &               4588968 \\
 5  &               50363136 &             351263232 \\
 6  &             3652143480 &           30873601080 \\
 7  &           307454670000 &         3082065903648 \\
 8  &         29310549057000 &       343320789071016 \\
 9  &       3133368921937824 &     42320100429654912 \\
10  &     370060173560963304 &   5709664512091086984 \\
11  &   47968071364509850944 & 837942419330764322976 \\
12  & 6756542767252059234840 &                       \\
\br
\end{tabular}
\end{indented}
\end{table}

\begin{table}
\caption{\label{tab2}Series for $\chi$ and $\chi_s$ for spin-$\frac{3}{2}$.
To avoid fractions a multiplier
$2^{n+2}n!/5$ ($2^{n+3}(n+1)!/5$) has been used for $\chi$ ($\chi_s$) series, where
$n$ is the power of $K$.}
\begin{indented}
\item[]\begin{tabular}{@{}rrr}
\br
n & $\chi$ & $\chi_s$ \\
\mr
\multicolumn{3}{c}{Simple Cubic Lattice $S=\frac{3}{2}$}  \\
 0  &                 1  &                  2  \\
 1  &                60  &                 60  \\
 2  &              1440  &               2220  \\
 3  &             50136  &             106032  \\
 4  &           2241660  &            6103230  \\
 5  &         124125372  &          417121164  \\
 6  &        8102868414  &        32715943017  \\
 7  &      613292153184  &      2911926450048  \\
 8  &    52599376466556  &    289263779556198  \\
 9  &  5056198898505288  &  31792485934519488  \\
\mr
\multicolumn{3}{c}{Body Centred Cubic Lattice $S=\frac{3}{2}$}  \\
 0  &                  1  &                   2  \\
 1  &                 80  &                  80  \\
 2  &               2720  &                4160  \\
 3  &             136448  &              283776  \\
 4  &            8751600  &            23240440  \\
 5  &          696028496  &          2263139152  \\
 6  &        65331028472  &        253095247076  \\
 7  &      7121212898544  &      32175304799424  \\
 8  &    879298191968624  &    4563926306507096  \\
 9  & 121768840349153216  &  716734730963510496  \\
\br
\end{tabular}
\end{indented}
\end{table}

\begin{table}
\caption{\label{tab3}Estimates  of critical point $K_C$ and exponent $\gamma$ (in
brackets) from $[N,D]$ Pad\'e approximants to the spin-$\frac{1}{2}$ uniform/staggered susceptibility
series. Defective PA's are denoted *.}
\begin{indented}
\item[]\begin{tabular}{ccccc}
\br
& \multicolumn{4}{c}{Spin $S=\frac{1}{2}$ } \\
 \mr
& \multicolumn{2}{c}{Simple Cubic} & \multicolumn{2}{c}{Body Centred Cubic} \\
\mr
$[N,D]$ & F ($\chi$) & AF ($\chi_s$) & F ($\chi$) & AF ($\chi_s$) \\
\mr
$[6,7]$ & 1.1900  & 1.0577  & 0.7935  & 0.7266  \\
      & (1.414) & (1.426) & (1.416) & (1.436) \\
$[7,6]$ & 1.1925  & 1.0611  & 0.7935  & 0.7266  \\
      & (1.432) & (1.455) & (1.416) & (1.435) \\
$[5,7]$ & 1.1914  & 1.0598  & 0.7937  & 0.7266  \\
      & (1.421) & (1.440) & (1.419) & (1.434) \\
$[6,6]$ & 1.1914  & 1.0597  & 0.7936  & 0.7264  \\
      & (1.421) & (1.439) & (1.417) & (1.431) \\
$[7,5]$ & 1.1931  & *       & 0.7939  & 0.7267  \\
      & (1.438)   &        & (1.423) & (1.436) \\
$[5,6]$ & 1.1910  & 1.0592  & 0.7937  & 0.7264  \\
      & (1.418) & (1.434) & (1.418) & (1.432) \\
$[6,5]$ & 1.1901  & 1.0583  & 0.7936  & 0.7264  \\
      & (1.411) & (1.425) & (1.418) & (1.432) \\
\br
\end{tabular}
\end{indented}
\end{table}

\begin{table}
\caption{\label{tab4}Estimates  of critical point $K_C$ and exponent $\gamma$ (in
brackets) from $[N,D]$ Pad\'e approximants to the Spin-1 uniform/staggered susceptibility
series. Defective PA's are denoted *.}
\begin{indented}
\item[]\begin{tabular}{@{}ccccc}
\br
& \multicolumn{4}{c}{Spin $S=1$ } \\
 \mr
 & \multicolumn{2}{c}{Simple Cubic} & \multicolumn{2}{c}{Body Centred Cubic} \\
\mr
$[N,D]$ & F ($\chi$) & AF ($\chi_s$) & F ($\chi$) & AF ($\chi_s$) \\
\mr
$[5,6]$ & 0.38478  &       & 0.26400  &   \\
      & (1.409)  &       & (1.404)  &   \\
$[6,5]$ & 0.38478  &       & 0.26398  &   \\
      & (1.409)  &       & (1.403)  &   \\
$[4,6]$ & 0.38478  & *     & 0.26397  & 0.25431  \\
      & (1.409)  &       & (1.403)  & (1.405) \\
$[5,5]$ & 0.38475  & *     & 0.26389  & 0.25431  \\
      & (1.408)  &       & (1.398)  & (1.401) \\
$[6,4]$ & 0.38467  & 0.36541 & *      & 0.25410 \\
      & (1.406) & (1.409)  &        & (1.395) \\
$[4,5]$ & 0.38487 & 0.36566  &  *     & 0.25410 \\
      & (1.411) & (1.417)  &        & (1.395) \\
$[5,4]$ & 0.38483 & 0.36565  & *      & 0.25396 \\
      & (1.410) & (1.417) &         & (1.390) \\
\br
\end{tabular}
\end{indented}
\end{table}

\begin{table}
\caption{\label{tab5}Estimates of the critical temperatures and leading susceptibility amplitudes,
from Pad\'e approximant analysis.}
\begin{indented}
\item[]\begin{tabular}{@{}cccccccc}
\br
\multicolumn{8}{c}{SC Lattice } \\
 \mr
 & \multicolumn{2}{c}{$S=\frac{1}{2}$} & \multicolumn{2}{c}{$S=1$} & \multicolumn{2}{c}{$S=\frac{3}{2}$}  & \multicolumn{1}{c}{$S=\infty$} \\
\mr
      & F ($\chi$)   & AF ($\chi_s$)  & F ($\chi$)    & AF ($\chi_s$)  & F ($\chi$)    & AF ($\chi_s$) & $\chi$ \\
\mr
$K_C$ & 1.192(2) & 1.059(2)  & 0.38478(15)     & 0.3656(2) & 0.195(3) &  0.190(1) & 0.69304  \\
$k_B T_c/J$ & 0.839(1) & 0.944(2) & 2.599(1) & 2.735(1)  & 5.13(8)  &  5.26(3)  &           \\
$k_BT_c/\tilde{J}$&1.119(2)&1.259(2)&1.2994(5)&1.3676(7) & 1.37(2)  & 1.404(7) & 1.4429  \\
$C_0$   &  1.26(2) & 1.20(3) & 1.11(2)        & 1.07(4)  &   &   & 0.9030  \\
\br
\multicolumn{8}{c}{BCC Lattice } \\
 \mr
 & \multicolumn{2}{c}{$S=\frac{1}{2}$} & \multicolumn{2}{c}{$S=1$} & \multicolumn{2}{c}{$S=\frac{3}{2}$} & \multicolumn{1}{c}{$S=\infty$} \\
\mr
      & F ($\chi$)   & AF ($\chi_s$)  & F ($\chi$)    & AF ($\chi_s$) & F ($\chi$)    & AF ($\chi_s$) & $\chi$ \\
\mr
$K_C$ & 0.7935(3) & 0.7266(2)  & 0.2640(2)  & 0.2542(2)      & 0.1354(10)  & 0.1327(4) & 0.48680   \\
$k_B T_c/J$ & 1.2602(5)  & 1.376(4)  & 3.788(2)  & 3.934(3)   &  7.39(5) & 7.54(2) &           \\
$k_BT_c/\tilde{J}$&1.6803(6)& 1.8350(5)& 1.894(1) & 1.967(1) & 1.97(1) & 2.009(6) & 2.0542  \\
$C_0$   &  1.15(2) & 1.10(3) & 0.98(1) & 0.94(1)             &  &  & 0.794 \\
\br
\end{tabular}
\end{indented}
\end{table}

\begin{table}
\caption{\label{tab6}Estimates of the secondary singularity, at $-K_N$ for the
uniform susceptibility and at $-K_C$ for the staggered susceptibility, for the
$S=\frac{1}{2}$ models on the BCC lattice.}
\begin{indented}
\item[]\begin{tabular}{@{}ccccc}
\br
$[N,D]$ & \multicolumn{2}{c}{F(K)} & \multicolumn{2}{c}{$F_s(K)$} \\
        & $K_C=0.7936$ & $\gamma=1.416$ & $K_C=0.7266$ & $\gamma=1.435$ \\
        & $-K_N({\rm est.}) $ & residue &  $-K_C({\rm est.}) $ & residue \\
$[5,7]$ & $-$0.7160  & 0.250 &  $-$0.7993 & 0.262  \\
$[6,6]$ & $-$0.7159  & 0.250 &  $-$0.7985 & 0.260  \\
$[7,5]$ & $-$0.7392  & 0.321 &  $-$0.8069 & 0.284  \\
$[5,6]$ & $-$0.7189  & 0.254 &  $-$0.7987 & 0.261  \\
$[6,5]$ & *         &        &  $-$0.7988 & 0.261  \\
$[4,6]$ & $-$0.7111  & 0.242 &  $-$0.7994 & 0.262  \\
$[5,5]$ & $-$0.7107  & 0.241 &  $-$0.7985 & 0.266  \\
$[6,4]$ & $-$0.7244  & 0.275 &  $-$0.7911 & 0.241  \\
\br
\end{tabular}
\end{indented}
\end{table}

\begin{figure}[htb]
  { \centering
    \includegraphics[width=0.45\columnwidth]{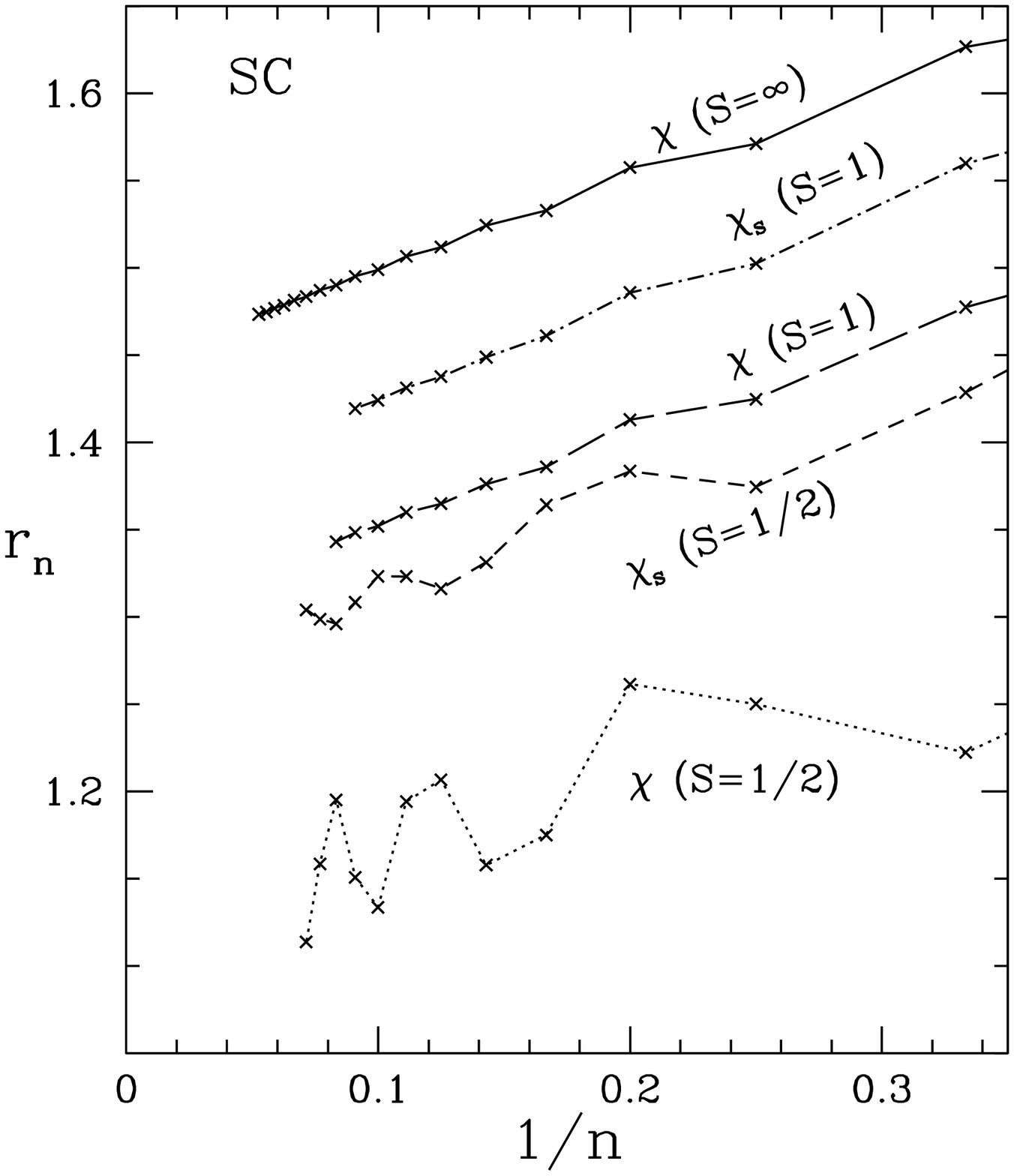}~
    \includegraphics[width=0.45\columnwidth]{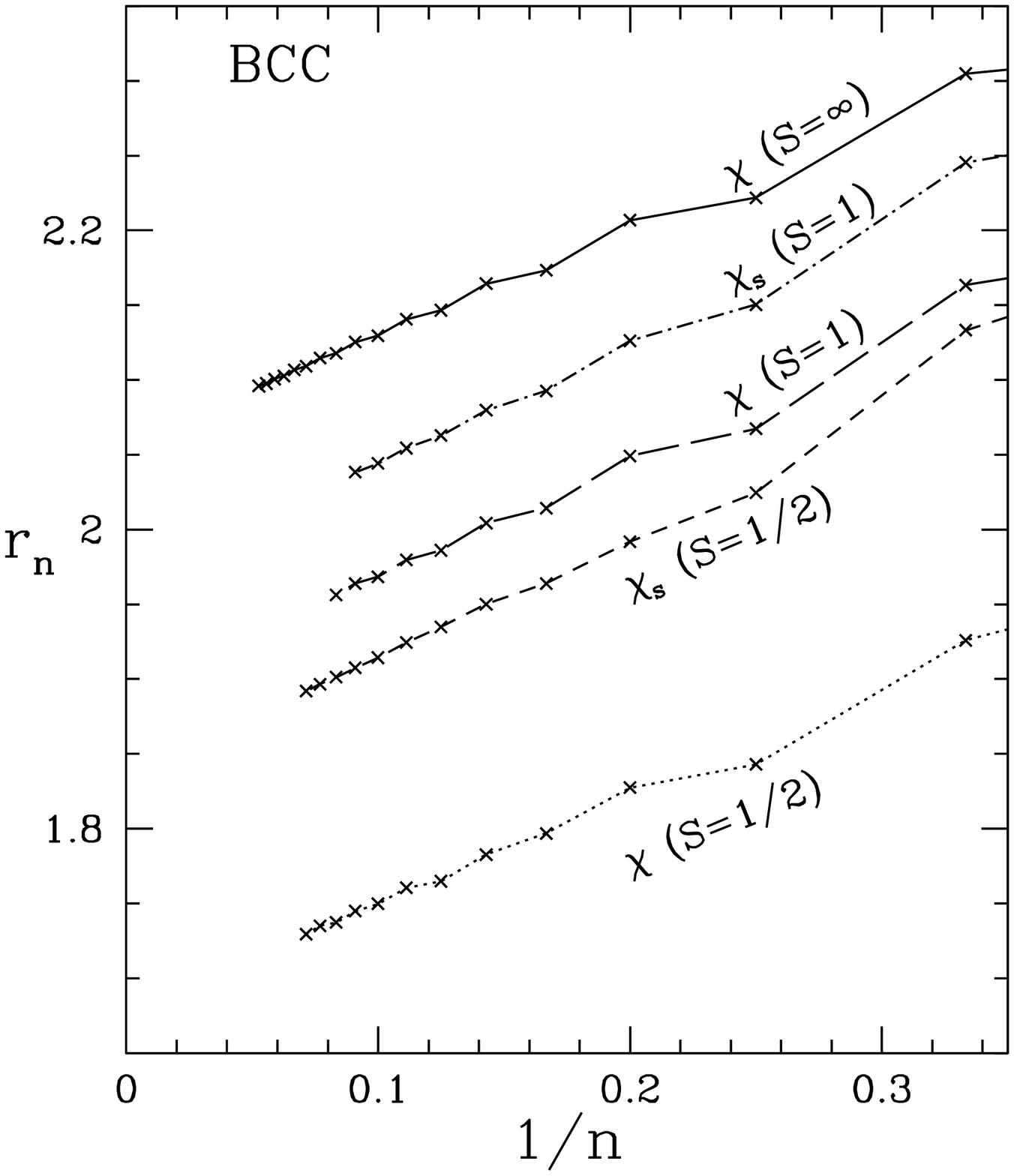}
  \caption{Ratio plots for the uniform and staggered susceptibilities for the SC and
  BCC lattices (as indicated) for $S=\frac{1}{2}, 1 , \infty$. The ratios are
  defined for the series in the variables $\tilde{K}=JS(S+1)/k_BT$, or equivalently
  $r_n = a_n/(S(S+1)a_{n-1})$ where $a_n$ are the coefficients of the
  $K$-series (\ref{eq_chi_ser}).}
}
\end{figure}

\begin{figure}[htb]
  { \centering
    \includegraphics[width=0.45\columnwidth]{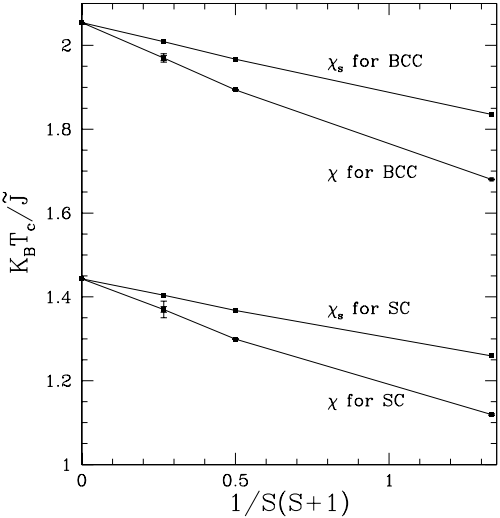}
  \caption{$k_B T_c/\tilde{J}$ versus $1/S(S+1)$.}
}
\end{figure}

\end{document}